# Emergent giant ferroelectric properties in cost-effective raw zirconia dioxide

Xianglong Li[1,5], Zengxu Xu[1,5], Songbai Hu[2,1,5], Mingqiang Gu[1], Yuanmin Zhu[3],

Qi Liu[1], Yihao Yang[1], Mao Ye[4], and Lang Chen[1*]

[1]Department of Physics, Southern University of Science and Technology, Shenzhen 518055, China.

[2]School of Physical Science, Great Basy University, Dongguan, 523429, China

[3]School of Materials Science and Engineering, Dongguan University of Technology, Dongguan 523808, China.

[4]School of Physics and Materials Science, Guangzhou University, Guangzhou, 510006, China.

[5]These authors contributed equally: Xianglong Li, Zengxu Xu and Songbai Hu.
[*]Corresponding author. Email: chenlang@sustech.edu.cn

## Abstract

Ferroelectric fluorite dioxides like hafnium ($HfO_2$)-based materials are considered to be one of the most potential candidates for nowadays large-scale integrated-circuits (ICs). While zirconia ($ZrO_2$)-based fluorites materials, which has the same structure as $HfO_2$ and **more abundant resources and lower cost of raw materials**, is usually thought to be anti- or ferroelectric-like. Here we reported a giant ferroelectric remnant polarization ($P_r$) amounted to 53 μC/cm² in orthorhombic $ZrO_2$ thin film at room temperature. This ferroelectricity arises from an electric field induced anti-ferroelectric to ferroelectric phase transition which is particularly noticeable at 77 K. Our work reveals the intrinsic ferroelectricity in $ZrO_2$ thin films and offers a new pathway to understand the ferroelectricity origin in fluorite oxides.

## One Sentence Summary:

Emergent giant $P_r$ ~53μC/cm² stemmed from irreversible phase transition driven by electric field in abundant $ZrO_2$.

## Main Text

The discovery of ferroelectricity in $HfO_2$-based fluorites thin films (*1*) has ignited widespread interests in exploring their applications in negative capacitance transistors, emerging memristor (*2-4*), neuromorphic computing (*5*) and artificial intelligence[6] because of its easy integrity with silicon-based semiconductor technologies. However, $ZrO_2$, being in the same fluorite family, was less that studied. This was probably due to that $ZrO_2$ usually appeared anti-ferroelectric or ferroelectric-like (*7-11*). **As a more plentiful and lower cost of raw materials compared to $HfO_2$, if it can achieve the same superior ferroelectric properties with $HfO_2$, it could greatly reduce production costs with the same manufacturing process to obtain ferroelectric properties.** Previous study reported that the ALD-deposited $ZrO_2$ thin films could be switched to anti-ferroelectric by electric field. (*7*) Anti-ferroelectric $ZrO_2$ could change to ferroelectric-like $ZrO_2$ by decreasing the film thickness. (*8*) PLD-deposited $ZrO_2$ also showed week-ferroelectricity through growth condition control. (*10, 11*) However, the ferroelectric performance of $ZrO_2$ is still far from the application. In addition, $ZrO_2$ has five different structures, which are non-polar monoclinic (*m*-, $P2_1/c$), tetragonal (*t*-, $P4_2/nmc$), orthorhombic anti-ferroelectric (*o-AFE*, Pbca) phases, and polar rhombohedral (*r-*, *R3m*), orthorhombic ferroelectric (*o-FE*, $Pca2_1$) phases. (*12-15*) These structures have very similar lattice parameters and energy levels which makes it difficult to tell them apart from each other. (*7, 15-20*)

The core issue in $ZrO_2$ thereby comes as to achieve a superior ferroelectric polarization and identify the corresponding crystal structure. Although the proximity of crystal structure, particularly between *t*-phase and o-*AFE* (and o-*FE*) phase, makes it hard to distinguish the ferroelectric parent phase, (*7, 15-20*) it provides possibility to harvest pure ferroelectric $ZrO_2$ through phase engineering. Previous studies have offered paths like elementary doping or substrate selection to obtain a relative stable ferroelectric o-*FE* phase. (*20, 21*) A recent study suggested that *t*-$ZrO_2$ would change to orthorhombic $ZrO_2$ (including *o-AFE*, *o-FE*) through a two-step phase transition during the growth. (*22*) These results implied that the acquiring of ferroelectric $ZrO_2$ was not that easy as it was sensitive to growth condition or external stimuli.

In this study, phase-pure *o*-$ZrO_2$ thin films with highly-improved crystallinity were deposited on (110) $SrTiO_3$ substrate (STO) using pulsed laser deposition (PLD). By carefully tailoring the growth window and a following electric activation, a giant remanent polarization ($P_r$) amounted to 53 μC/cm$^2$ was observed at room temperature. The structure of $ZrO_2$ thin films was thoroughly investigated by X-ray diffraction (XRD) and scanning transmission electron microscopy (STEM). It reveals that this giant $P_r$ arises solely from orthorhombic $ZrO_2$. Both experiment and density function theory (DFT) calculations imply that this giant $P_r$ stems from the electric field activated anti-ferroelectric-to-ferroelectric (*AFE*-to-*FE*) phase transition, which converts those as-grown anti-ferroelectric and ferroelectric blends into purely ferroelectric $ZrO_2$.

$ZrO_2$ thin films were grown on $La_{0.67}Sr_{0.33}MnO_3$ (LSMO ~ 25 nm) buffered STO by PLD. XRD *2θ-ω* scans for different growth conditions are shown in fig. S1A, B, and C. The optimal growth conditions for $ZrO_2$ thin films with the best crystalline quality and ferroelectric polarization was a temperature of 850°C, oxygen pressure 10 Pa, and thickness 12 nm (fig. S1D-F and fig. S4B-D). The thickness of each layer is confirmed by STEM cross-section in fig. S1g. The crystallinity under different growth conditions was determined by comparing the full width at half maximum (FWHM) from XRD rocking curves.

Then we grew $ZrO_2$ on (110) and (001) STO to find out a suitable substrate orientation, respectively. The XRD 2theta scans in Fig. 1A discloses a single peak at 30.3° on (110) STO in addition to LSMO, which is attributed to pseudo-cubic $ZrO_2$ (111) plane (*15, 23-25*); while on (001) STO, an additional *m*-$ZrO_2$ (002) diffraction is observed. The step-terraces illustrated in inset of Fig. 1A reveal an atomically flat surface for films grown on (110) STO, with a root-mean-square roughness of approximately 0.25 nm. Fig. 1B compared the full width at half maximum (FWHM) of $ZrO_2$ (111) and found that $ZrO_2$ on STO (110) exhibit a significantly narrower FWHM (0.1°) than that on STO (110) (0.17°), indicating an improved crystallinity by orientation control. The thin films grown on (110) STO thereby exhibit enhanced ferroelectric polarization (fig. S4E), which is consistent with previous reports. (*20*) Figure 1C investigated the thermo-stability of $ZrO_2$ by carrying out a temperature-dependent X-ray diffraction. It is observed that there is no significant change in crystal structure of the thin films throughout heating up to the growth temperature of 850°C and then cooling down to room temperature.

Figure 1D displays the XRD pole figure at $\chi$ =71° for ZrO$_2$ (111) on STO (110) substrate. The result reveals a six-fold crystal geometry in out of plane phase structure, where little shift was observed away from the red dashed circle. This result is confirmed again in Fig. 1E by a more detailed *2θ* scan ($\chi$ =71°) at different *φ*-angles, which could sufficiently exclude the presence of *r*- ZrO$_2$ in our thin films. (*11, 25*) Figure 1D also reveals an obvious peak-splitting (~7°) of ZrO$_2$ {111} diffractions. Such paired-spots, which are observed in Fig. 2A and fig. S2A, B &C as well, reflect an in-plane microstructure rotation in ZrO$_2$ thin film. A HR-STEM image in fig. S2A implies that this paired-spots actually originate from a twinning structure in ZrO$_2$/LSMO/STO (110) sample. In order to confirm the lattice parameters of ferroelectric ZrO$_2$, an in-plane *2θ-ω* scan is performed for the ZrO$_2$/LSMO/STO (110) sample in Fig. 1F. Except the diffraction peak at ~50.6° which was attributed to ZrO$_2$ {0-22} no other peak was detected. In-plane *2θ* scans for {-220} at different *φ*-angles in fig. S2C demonstrating little peak deviation from each other shown in fig. S2D. The distinct diffraction peaks in XRD patterns, resulting from subtle variations in interplanar-spacing due to different lattice constant, exhibit a diffraction distribution from the pattern of *t*-ZrO$_2$ (*10*), thereby allowing *o*-ZrO$_2$ to be distinguished from *t*-ZrO$_2$. Then a further wide-range reciprocal space mapping (WARSM) in Fig. 2A discloses another peak at 43°, which arises from the ZrO$_2$ (1-21) diffraction. Based on the three angles, the lattice parameters of ZrO$_2$ are calculated under either orthorhombic or tetragonal geometry, which turns out to be *a*=5.096 Å, *b*=5.171 Å, *c*=5.057 Å or *a*=*b*=5.097 Å, *c*=5.128 Å, respectively. As the former one is quite close to the typical *o-FE* structure, while the latter deviates significantly from the reported tetragonal structure (*10, 12-14*), the as-grown ZrO$_2$ is identified to have an orthorhombic structure.

The epitaxial relationship between ZrO$_2$, LSMO buffer layer and STO substrate are investigated by an in-plane WARSM in Fig. 2A. Besides those diffractions of LSMO and STO, no other spots but those belonging to *o*-ZrO$_2$ {022} and {112} were detected. The angle between (0-22) and (-1-12) is measured to be 30°. This value is very close to the theoretical angle, implying that both diffractions stem from *o*-ZrO$_2$ rather than impurity phases or textures. The epitaxial relationship between ZrO$_2$ and LSMO/STO is determined as ZrO$_2$ [0-22] // LSMO [001] // STO [001] and ZrO$_2$ [111] // LSMO [110] // STO [110] in in-plane and out-of-plane direction, respectively. The microscopic structure of the ZrO$_2$ thin film was investigated by high resolution (HR)-STEM. Fig. 2B and C display the high-angle annular dark field (HADDF)-STEM images of the ZrO$_2$/LSMO/STO (001) plane and (-110) plane, respectively. The Fast Fourier Transform (FFT) demonstrated a single *o*-phase in our thin films. In fig. S2A, a twinning structure manifested by a 37° small angle grain boundary was captured along the STO [001] direction. This microscopic structure actually resulted from the 7° rotation of the (0-22) plane measured in Fig. 2A. At the ZrO$_2$/LSMO interface, the Zr atoms did not follow the stacking manner of LSMO, but propagated in a domain-match-epitaxy (DME) mode. This DME stacking is evidenced by the filtered strip-structure in Fig. 2B & C, which is same to that of $HfO_2$-based thin films grown on LSMO (*24*). Due to the mismatch in the in-plane lattice parameters between $ZrO_2$ and LSMO, a strain is induced at their interface. To accommodate this strain and reduce the interfacial energy, the $ZrO_2$ [0-22] rotates a small angle (α/2 ~3.5º) away from the LSMO [001]. This rotation results in a specific matching relationship between the two materials, which is crucial for understanding their interaction and the overall behavior of the composite system. The schematic diagram of this matching relationship, highlighting the angle of rotation and the lattice misalignment, is illustrated in fig. S2E.

Ferroelectric properties of $ZrO_2$ thin films are shown in Fig. 3. Note that the leakage current is excluded by a *PUND* (positive up negative down) method shown in fig. S3A. A superior $P_r$ of 53 μC/cm² is measured in the films with optimal growth conditions compared to the other one, shown in Fig. 3A and fig. S3B-E. This value significantly exceeds those textured polymorph or $ZrO_2$ mixture, and is at the same level of the best $HfO_2$-based ferroelectric materials, see Fig. 3B. Similar to previous studies, our thin films also require an electric field to 'wake-up' the ferroelectricity. However, this 'wake-up' is different from other reports, as the former needs only one cycle pulse triangular wave (voltage curve see fig. S3A) to activate the ferroelectricity, while the latter requires to run many times of cycling. (*32-34*) In addition, the thin film could be only activated for voltage higher than 5 V (fig. S4A & B). For lower than 4.5 V, the thin film still demonstrated pinched hysteresis loops even after $10^5$ cycles at room temperature and $10^8$ cycles at 77 K (fig. S4C & D). These results indicate that the ferroelectricity was not obtained by training, but rather related to a mutation such as phase transition.

In Fig. 3A the leakage current could not be subtracted completely at room temperature as seeing the $P_r$ is not saturated at higher filed. To obtain the intrinsic ferroelectric polarization in our thin films, P-E loops by decreasing the temperature from 295 K to 77 K were measured. The results for pristine and 5 V activated $ZrO_2$ are shown in Fig. 3 C and D, respectively. The corresponding *P-E* and *I-E* loops with *PUND* method are shown in fig. S4E and F. As the temperature drops to 130 K, the leakage current peak disappears. The thin film demonstrates a typical ferroelectric polarization saturation at approximate ~ 20 μC/cm², at around 150 K. By reducing the temperature to 77 K step by step, the current decrease in further and the curves start to stick together at 130 K, 110 K, 90 K and 77 K, indicating that the leakage current has been excluded completely. The intrinsic $P_r$ was measured to be ~ 15 μC/cm² (fig. S4G).

In Fig. 3C, the pristine $ZrO_2$ displays a pinched anti-ferroelectric-like *P-E* loop which presents four switching charge peaks on *I-E* curves. Differently, the activated $ZrO_2$ in Fig. 3D exhibited a typical ferroelectric *P-E* loop and two switching charge peaks. Charge vs. electric field (*Q-E*) curves for pristine and activated sample are shown in fig. S4H, I, where the former one exhibits an anti-ferroelectric cross over at approximate 2 MV/cm, while latter one acts in a typical ferroelectric manner. The anti-ferroelectric-like *P-E* loop and *Q-E* curves in pristine ZrO$_2$ thin films may be attributed to the coexistence of o-*AFE* and o-*FE* phases. Under 5 V actuation at room temperature, the as-grown anti-ferroelectric and ferroelectric blends is able to transit to pure o-*FE*.

Fig. 3G - I demonstrate the electric polarization switching of ZrO$_2$ thin film through piezoelectric force microscope (PFM). Fig. 3G and H show the PFM poling map written at +5 V on the larger box and −5 V on the smaller box. A 180° phase contrast between the square domain patterns qualitatively demonstrates the appearance of the up- and down-polarization, respectively, which illustrates that a switchable polarization exists in the ZrO$_2$ thin film at room temperature. The single force pattern of PFM measurement is shown in Fig. 3I. The local out-of-plane piezo response of the ZrO$_2$ thin film was measured on the bare surface as a function of a voltage applied to a conductive tip. That amplitude hysteresis loop and butterfly phase curves show clearly the symmetric ferroelectric switching behavior, indicating the bi-stability of polarization states in our heterostructures. The amplitude is not perfectly symmetric across the domain walls, probably due to long-range electrostatic interactions between the conductive tip and the sample, whicFh is always traced from ferroelectric thin films in PFM measurement.

The room temperature fatigue test for the ZrO$_2$ thin film under 6 V is obtained in figure S5A. The *P-E* loops before and after the test are shown in fig. S5B. P$_r$ ~ 20 µC/cm$^2$ is still preserved after 10$^7$ cycles. Although this P$_r$ decrease a lot, it is comparable to those HfO$_2$-based thin film and better than other reported ZrO$_2$ (see Fig. 3B). The leakage current density curves of ZrO$_2$ thin film are measured between 4.0 ~ 7.0 V, displayed in figure S5C.

In order to understand the activation behavior and the giant ferroelectric polarization in ZrO$_2$ thin films, the free energy of the both *o-FE* and *o-AFE* phases is calculated by DFT calculation. The atomistic structures are shown in Fig. 4A. The two structures show similar arrangements of oxygen and Zr atoms, which makes them indistinguishable from each other by using XRD or our STEM. We calculated the total energy versus strain along [-211] and [111] orientations (with fixed strain along the other two directions) in Fig. 4B and C. The strain of the above orientation was computed at table S1. The two *o*-ZrO$_2$ structures remain at very close energy scales over the strain span, which means *o-AFE* could easily transit to *o-FE* upon external stimuli like electric filed. Similar phenomena have been reported in HfO$_2$-based thin films (*9, 13, 14*). This result well supported the activation behavior described in Fig. 3. We also predicted the evolution of ferroelectric polarization for the *o-FE* along the polarization flipping path and the corresponding energy profile in the thin films in fig. S6. The theoretical predicted ferroelectric polarization value is ~ 66 µC/cm$^2$, which is similar to the result of ref. 35 & 36 and closes to our experiment value.

## Conclusions

In this study, high-quality phase-pure $ZrO_2$ films were grown on LSMO buffered (110) STO by PLD. XRD and HR-STEM analysis identified that the as-grown $ZrO_2$ films have an orthorhombic structure and consist of in-plane twinning crystal. A ferroelectric polarization of 53 μC/cm² was obtained in activated $ZrO_2$ at room temperature. Both experiment and theoretical calculated supports that this superior ferroelectric polarization stems from the electric field induced *o-AFE* to *o-FE* phase transition, which converts those as-grown anti-ferroelectric and ferroelectric blends into purely ferroelectric $ZrO_2$. Our study reverses the poor ferroelectric opinion on $ZrO_2$ by exhibiting a superior $P_r$ which is on par with $HfO_2$ and most $HfO_2$-based thin films and open the door of using ferroelectric $ZrO_2$ thin films in microelectronics.

**Acknowledgments:**
This work was supported by the National Key R&D Program of China (No. 2022YFA1402903)
The Guangdong Provincial Key Laboratory Program (Grant No. 2021B1212040001)
Science Technology and Innovation Commission of Shenzhen Municipality (JCYJ20210324104812034)
Guangdong Innovative and Entrepreneurial Research Team Program (Grant No. 2021ZT09C296)
Natural Science Foundation of Guangdong Province (2021A1515110389)
SUSTech Core Research Facilities.

**Author contributions:**
L. C. conceived the idea and designed the research, X. L. L. & Z. X. X. carried out the experiments, Y. H. Y., Q. L. & M. Y. analyzed the results, X. L. L., Z. X. X. & S. B. H. wrote the paper. M. Q. G. carried out the DFT calculations. X. L. L. & S. B. H. assist in carrying out the structure characterization. X. L. L. & Z. X. X. assist in carrying out the ferroelectric characterization. S. B. H. assist in the sample fabrication. L. C. revised the manuscript.

**Competing interests:**
Authors declare that they have no competing interests.

**Data and materials availability:**
The data that support the findings of this study are available from the corresponding author upon reasonable request.

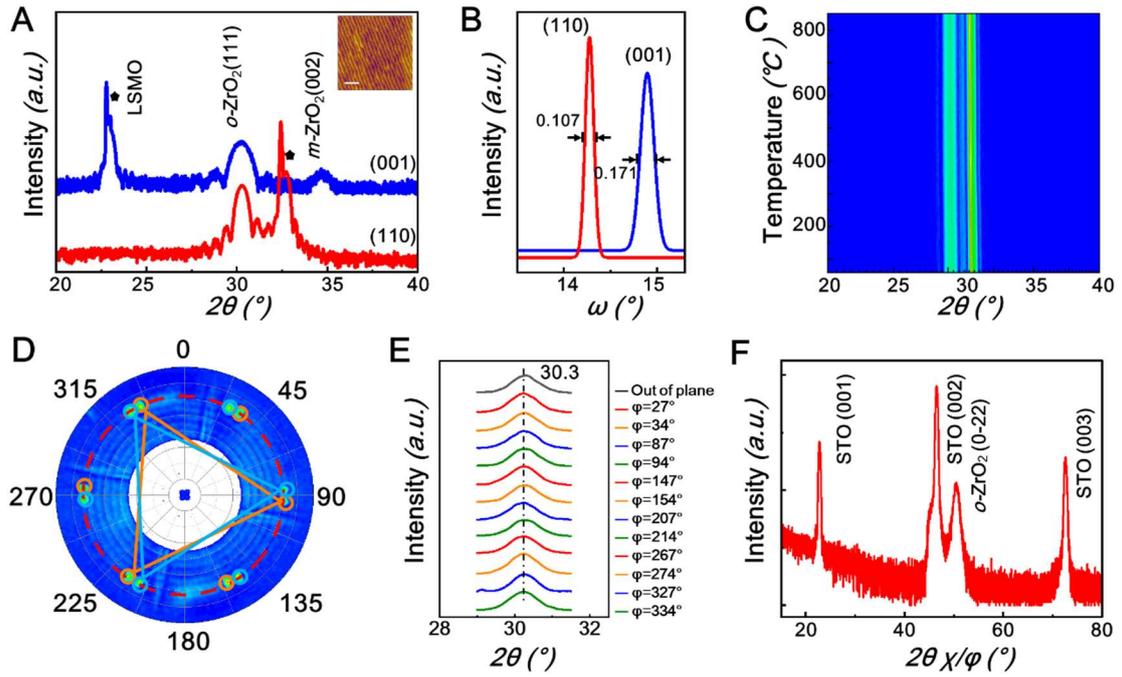

**Fig. 1. Structural characterization of ZrO₂/LSMO heterostructures.** (**A**) XRD *2θ–ω* scan of ZrO₂/LSMO bilayer grown on (110), (001) STO. The illustration is an Atomic Force Microscopy (AFM) topography characterization for ZrO₂/LSMO bilayer grown on (110) STO, of which scale bar is 2 μm. The peaks of LSMO in (**A**) are marked by black stars. Rocking curve (**B**) and variable-temperature X-ray diffraction (**C**) for ZrO₂/LSMO on STO (110) substrates, temperature range from 60°C to 850°C. (**D**) Pole Figure of ZrO₂ on STO (110). The red dashed circle dictates *χ*=71°. (**E**) *2θ* scan at *χ*=71° over different *φ*. The black curve represents the out-of-plane scan at *χ*=0°. (**F**) In-plane 2θ *χ/φ* scan for ZrO₂/LSMO (110)/STO (110) sample.

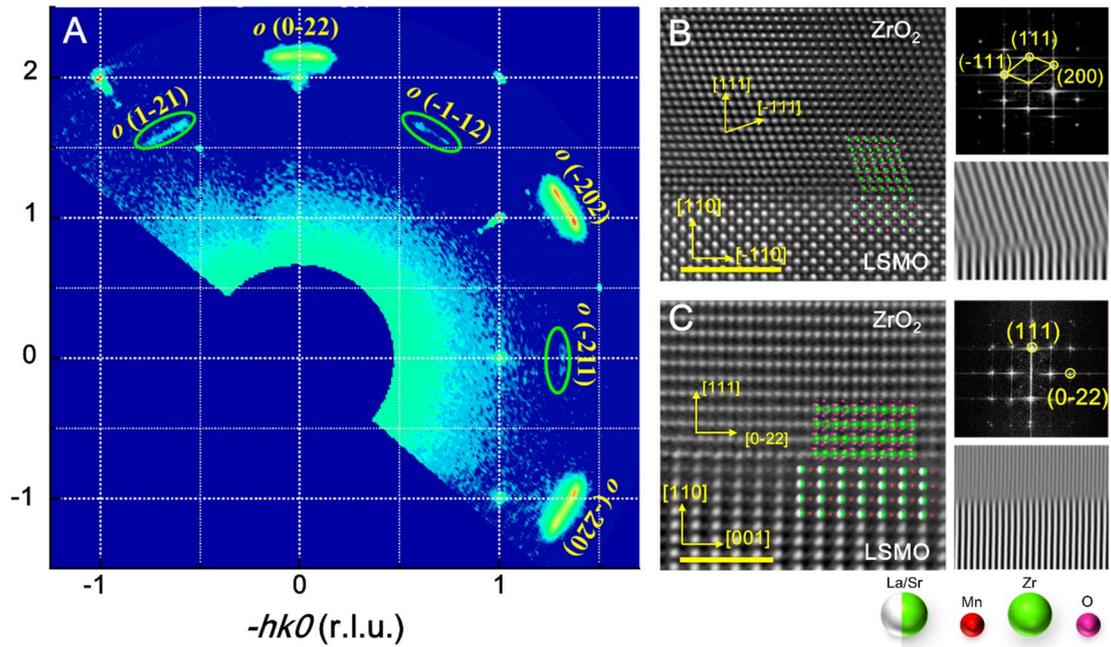

**Fig. 2. Epitaxial relationship of the heterostructures.** (**A**) In-plane WARSM for $ZrO_2$/LSMO bilayer grown on STO (110). The white grids corresponding to integers -*hk0* and *00l* are guides to eyes. The yellow-colored index represents diffraction belonging to orthorhombic $ZrO_2$. The STEM-HADDF images of the $ZrO_2$/LSMO/STO (001) plane and (110) plane are given in (**B**) and (**C**) respectively. The scale bar is 3nm. The Fast Fourier Transform (FFT) of $ZrO_2$ lattice and the epitaxial matching relationship between $ZrO_2$ and LSMO are shown at the right panel.

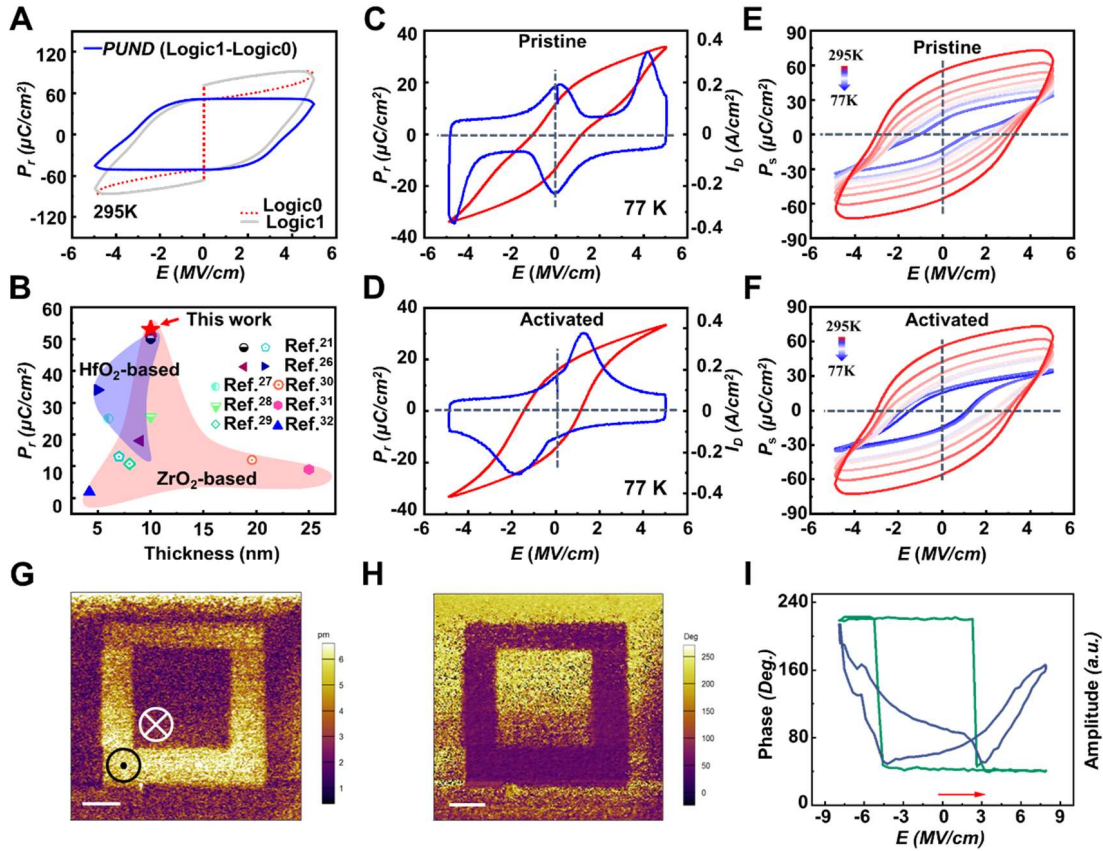

**Fig. 3. Ferroelectric properties for the ZrO₂ thin films.** *P-E* loops for Pt/ZrO₂/LSMO structure (**A**) at room temperature (**B**) Reported $P_r$ in HfO₂ or ZrO₂-based thin films.(*21, 25-32*) *P-E* and *I-E* for (**C**) pristine and 5.0 V (**D**) activated ZrO₂ respectively. The pristine (**E**) and (**F**) activated ZrO₂ by varying temperature under 5 *MV*/cm. The amplitude (**G**) and (**H**) phase of the PFM poling map written at +5 V on the larger box and -5 V on the smaller box. The scale bar is 1 μm. (**I**) The local hysteretic behavior of phase and amplitude tipped on the bare surface as a function of bias.

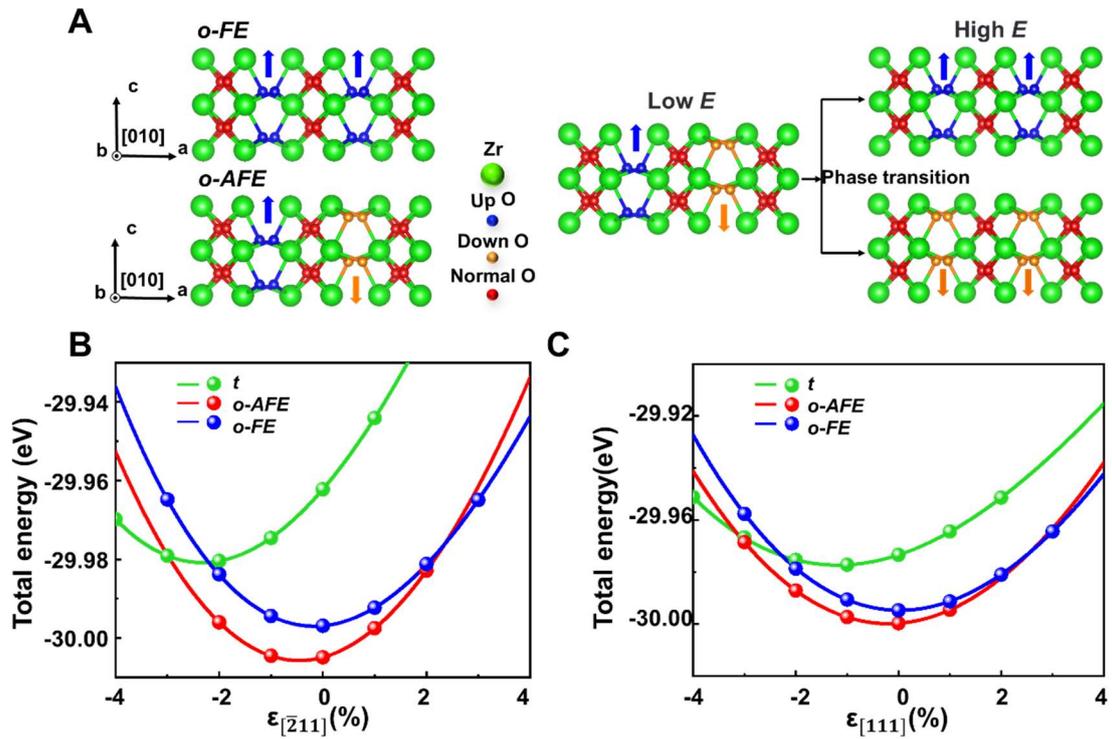

**Fig. 4. DFT calculations of phase energy levels in the ZrO₂ thin films.** (**A**) Atomistic structure of *o-FE* and *o-AFE* ZrO₂ and phase transition schematics from *o-AFE* ZrO₂ to *o-FE* ZrO₂ upon special electric field. Total energy versus strain along (**B**) [-211] orientation and (**C**) [111] orientation.

Supplementary Materials for

# Emergent giant ferroelectric properties in cost-effective raw zirconia dioxide

Xianglong Li[1,5], Zengxu Xu[1,5], Songbai Hu[2,1,5], Mingqiang Gu[1], Yuanmin Zhu[3],

Qi Liu[1], Yihao Yang[1], Mao Ye[4], and Lang Chen[1*]

[1]Department of Physics, Southern University of Science and Technology, Shenzhen 518055, China.

[2]School of Physical Science, Great Basy University, Dongguan, 523429, China

[3]School of Materials Science and Engineering, Dongguan University of Technology, Dongguan 523808, China.

[4]School of Physics and Materials Science, Guangzhou University, Guangzhou, 510006, China.

[5]These authors contributed equally: Xianglong Li, Zengxu Xu and Songbai Hu.
[*]Corresponding author. Email: chenlang@sustech.edu.cn

**This PDF file includes:**
Materials and Methods
Supplementary Text
Fig.S1 to S6
Table S1

## Materials and Methods

**Thin film growth and capacitors fabrication.** 12 nm $ZrO_2$ thin films were grown by PLD on LSMO (25 nm) buffered STO substrates with a KrF excimer laser (wavelength =248 nm). The optimum growth conditions were 750 °C, 100 mTorr $O_2$, 1.5 mJ/cm$^2$, 2 Hz and 800 °C, 100 mTorr $O_2$, 1.3 mJ/cm$^2$, 2 Hz for LSMO and $ZrO_2$ films, respectively. The bottom and top LSMO layers were deposited with the same growth condition. The $ZrO_2$/LSMO bilayer was cooled down to room temperature at a rate of 10 °C per minute under an oxygen pressure of 10 kPa. 20 nm of Pt were sputtered on LSMO/$ZrO_2$/LSMO/STO at room temperature to fabricate the MIM (metal-insulator-metal) capacitors.

**Physical and Electrical Characterization.** The structure of the samples was characterized by X-ray diffraction (XRD) using a Rigaku SmartLab diffractometer (Cu Kα radiation and λ =1.5406Å). The out-of-plane measurements were carried out in a high-resolution PB-Ge(220) × 2 scattering geometry. The in-plane measurements were done in a medium resolution PB scattering geometry with Cu K$_β$ absorbance equipped at the receiving end. The scan speed was 6°/min. The STEM samples were prepared using a focused ion beam milling system (FEI Helios NanoLab dual beam system). A double-aberration-corrected transmission electron microscope (Thermo Fisher Themis Z G2 60-300) equipped with a super EDS detector operated at an accelerating voltage of 300 kV was employed to obtain the STEM imaging and energy-dispersive X-ray spectroscopy (EDS). The probe convergence angle and the detection angle used for HAADF-STEM imaging were 25 and 64-200 mrad, respectively. The atomic STEM-EDS mapping was obtained with a beam current of ~0.15 nA for ~8 min. The surface topography and piezoelectric response of the as-grown $ZrO_2$/LSMO films samples were characterized by using a commercial scanning probe microscope (MFP-3D; Asylum Research) and Ir/Pt-coated atomic force microscopy tips (PPP-EFM; NanoWorld) at room temperature. The *P-E* and *I-E* loops of LSMO/$ZrO_2$/LSMO tri-layer were measured by a RADIANT TECHNOLOGIES PRECISION MULTIFERROIC system at 1 kHz at room temperature.

**Density functional theory (DFT) simulation.** DFT simulation was performed using a Vienna Ab-initio Simulation Package (VASP) code to understand the ferroelectric property and the energy profile of different phases of $ZrO_2$ under different strains. The projector augmented wave (PAW) method was used to treat the core and valence electrons. The revised Perdew-Burke-Ernzerhof (PBE) functionals for solids (PBEsol) were used in our calculation. A Γ-centered fine k-point grid with the resolution of 0.02 (in unit of 2π/Å) is used. For the structure optimization, the atomic positions were relaxed until the forces on each atom were less than 1 meV/Å. The ferroelectric polarization was computed using a Berry phase method.

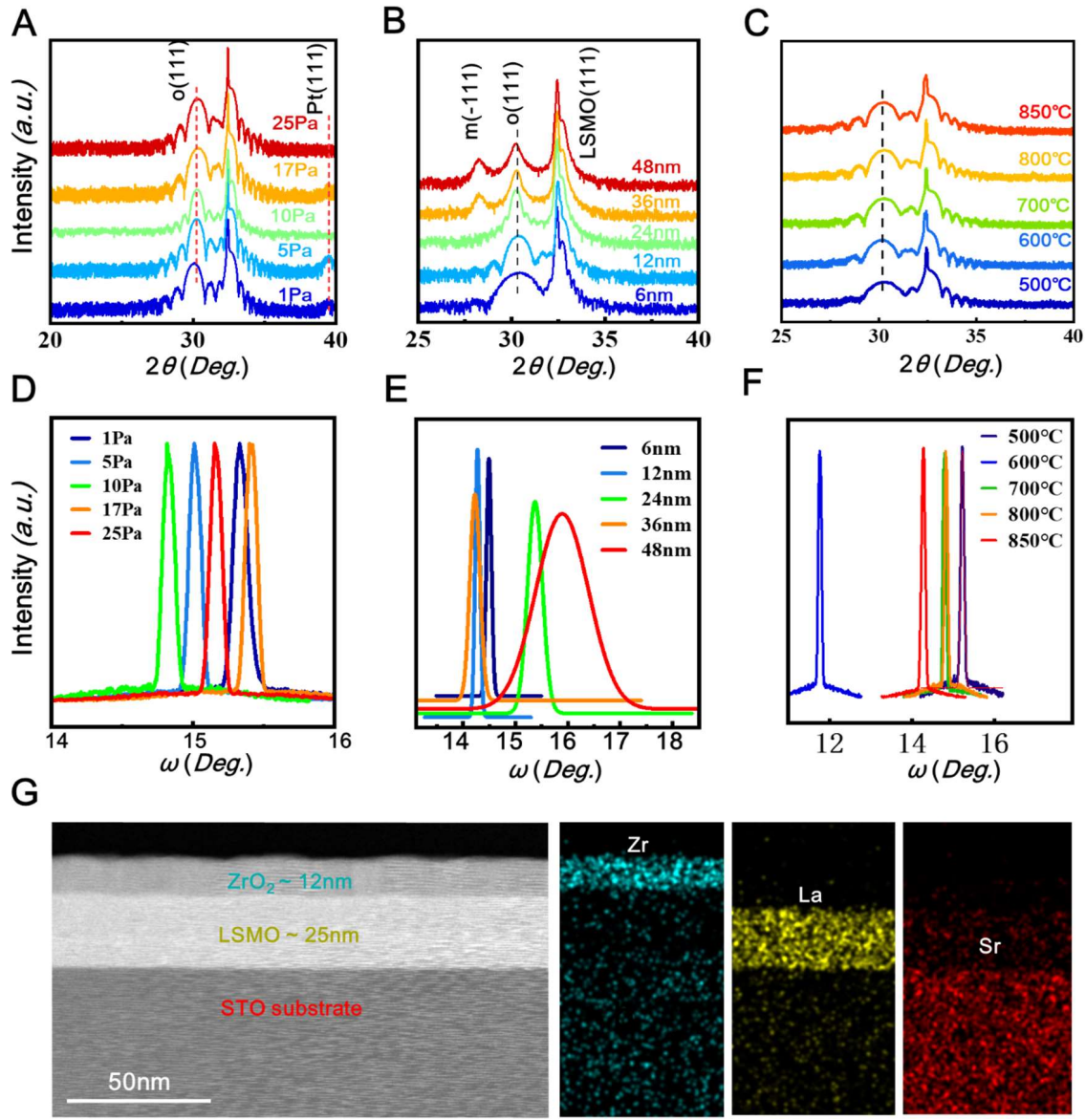

**Fig. S1. HR-STEM and XRD characterizations of ZrO$_2$ thin films.** The thickness of ZrO$_2$ and LSMO were measured to be 12nm and 25nm, respectively. The *2θ-ω* evolution of ZrO$_2$ thin films under different oxygen pressure, film thickness and temperature are shown in (**A**), (**B**), and (**C**) respectively. Rocking curves for ZrO$_2$ (111) at *2θ*=30.2° under different oxygen pressure, film thickness and temperature are shown in (**D**), (**E**) and (**F**) respectively. (**G**) Cross-sectional STEM and energy-dispersive X-ray spectroscopy mapping of ZrO$_2$/LSMO bilayer on (110) STO.

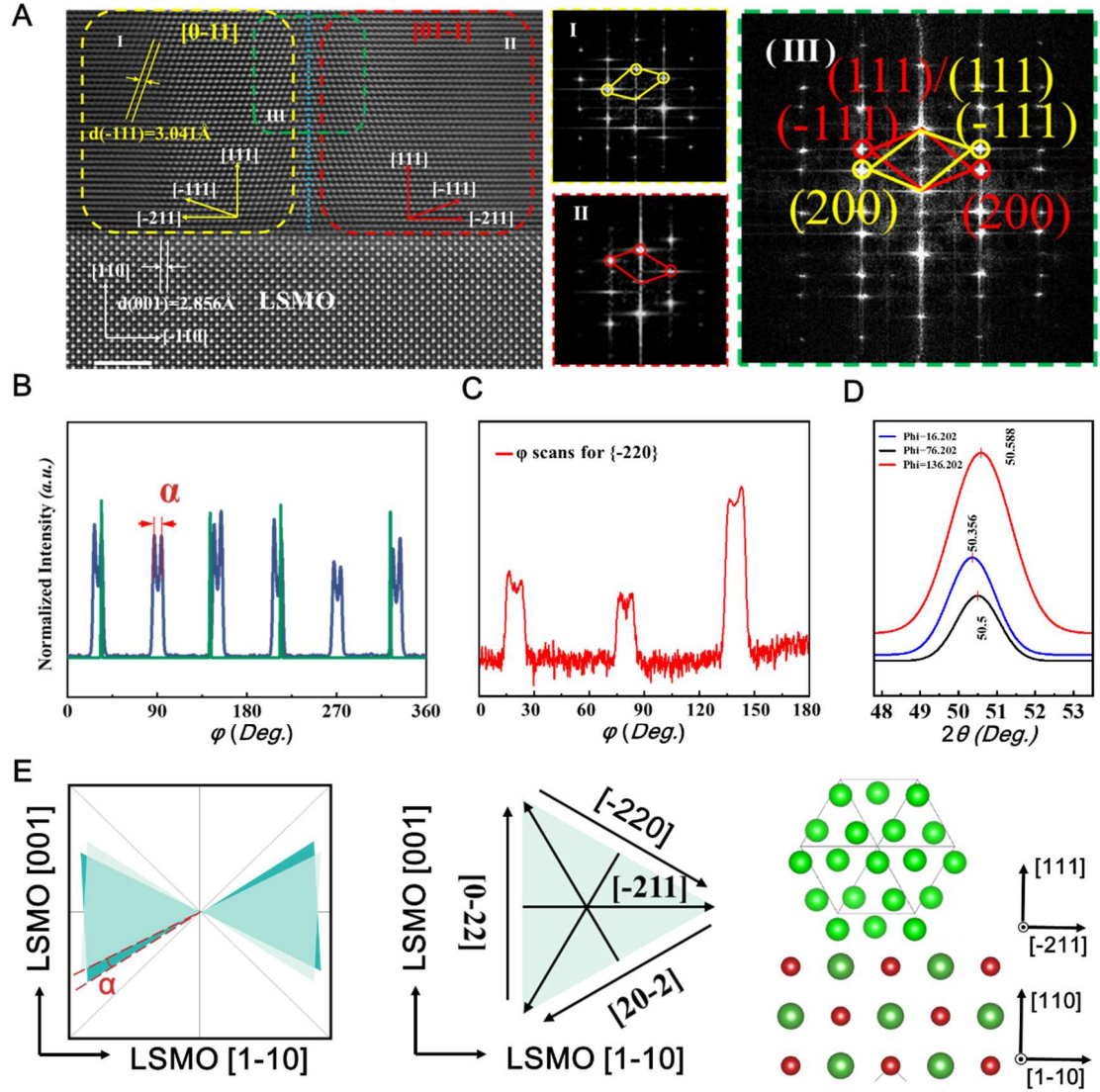

**Fig. S2. The structure and epitaxy matching relationship of ZrO$_2$ and LSMO bilayer.** (**A**) The STEM image in (**A**), reveals a twinning crystal structure for ZrO$_2$ on LSMO, evidenced by the FFT of regions I, II and III, where a 180º rotation geometry is observed. (**B**) Phi scans at *2θ*=32.84° and 30.3° for LSMO (110) and ZrO$_2$ (111), respectively. (**C**) Phi scans for {-220} at *2θ* = 50.6°. (**D**) *2θ* scan over different *φ*-angles and the peaks with the single peak smoothed in function of Gaussian. (**E**) The in-plane domain-match-epitaxy (DME) relationship between ZrO$_2$/LSMO. The angle α represents the in-plane rotation between two ZrO$_2$ flakes and is measured to be 7º according to (**D**)

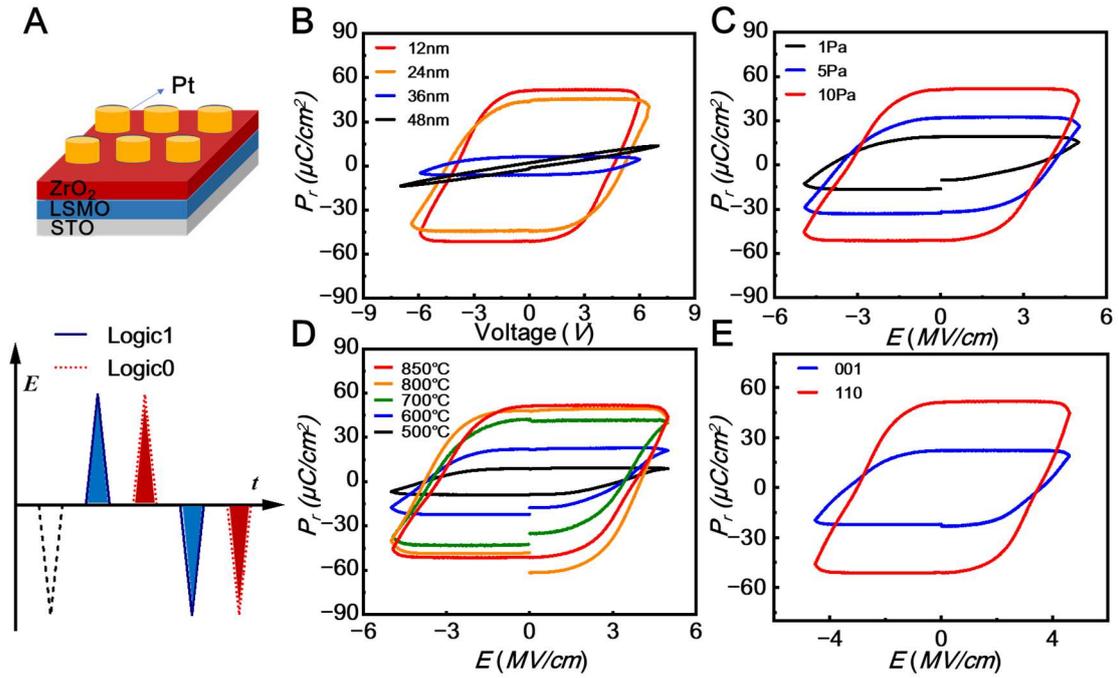

**Fig. S3. Ferroelectric characterizations on LSMO/ZrO$_2$/LSMO tri-layer capacitor under different conditions.** (**A**) The schematic capacitor device and two sequential wave functions (denoted as Logic 1 and 0) during the *PUND* measurement. (**B**) – (**E**), The *P-E* loops for various growth conditions during the *PUND* measurement.

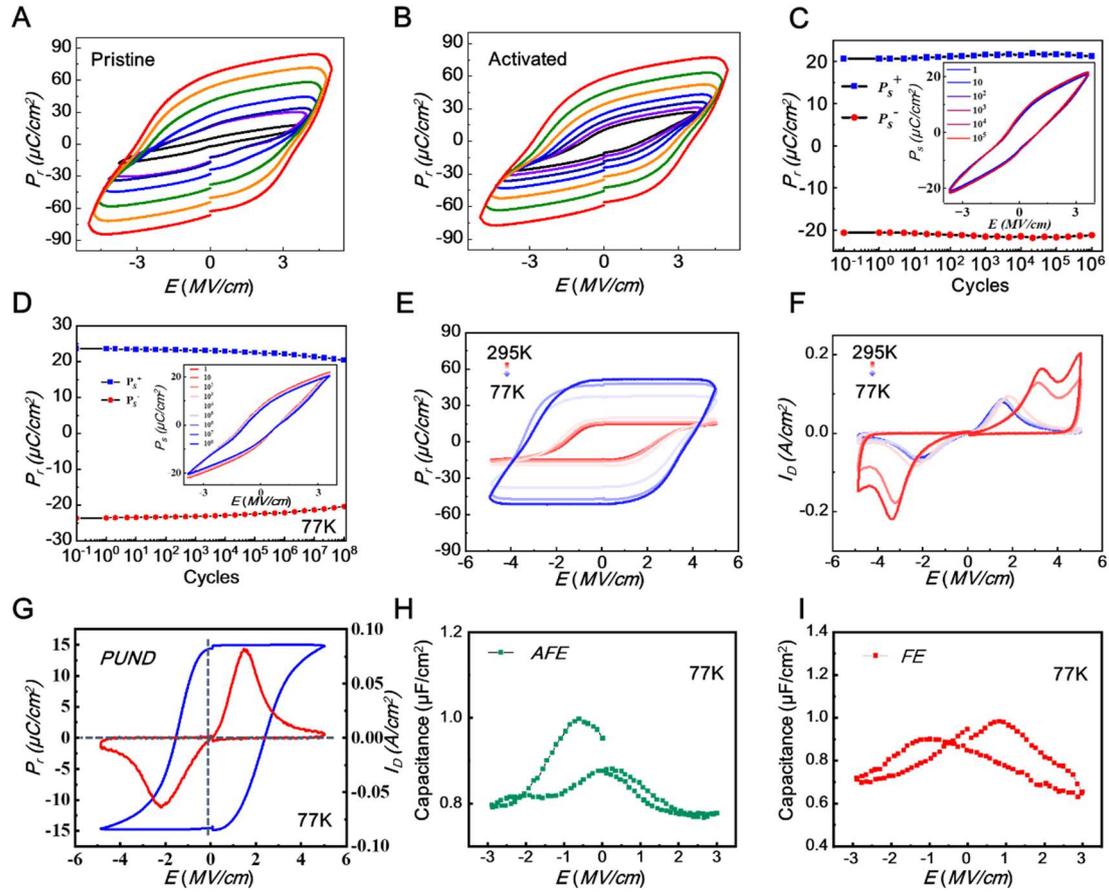

**Fig. S4. The transition from Anti-ferroelectric performance (pristine) to Ferroelectric performance (activated).** The measurement of *P-E* loops at room temperature by varying voltage between 4.5 ~ 6.0 V for (**A**), pristine and (**B**), activated thin films. The fatigue of Pt/ZrO$_2$/LSMO capacitor under 4.5V (**C**) at room temperature and (**D**) 77K respectively. The *P-E* and C-E loops with *PUND* measurement by decreasing temperature in activated thin films are displayed in (**E**), (**F**), (**G**), *P-E* and *I-E* loops with *PUND* measurement at 77K. Q-E curves for pristine (anti-ferroelectric state) and activated (ferroelectric state) sample is shown in (**H**) and (**I**) respectively.

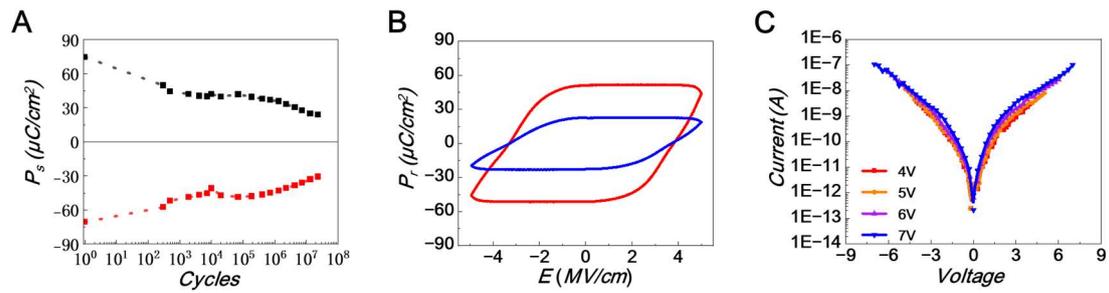

**Fig. S5. The measurements of fatigue and leakage properties in ZrO$_2$ thin films.** (**A**) The fatigue of Pt/ZrO$_2$/LSMO capacitor under 5*MV*/cm. (**B**) The *P-E* loops before and after the fatigue test. (**C**) The Leakage current density as a function of applied voltage for the capacitance measured between 4.0 ~ 7.0 V.

| lattice constant (Å) | a | b | c |
|---|---|---|---|
| Experiment | 5.1709 | 5.0961 | 5.0569 |
| o-FE Standard (7) | 5.2600 | 5.0680 | 5.0770 |
| o-AFE Standard (10) | 5.1869 | 5.2952 | 10.1675 |

| $ZrO_2$ o-FE structure | Experimental structure | Strains along different crystal orientations |
|---|---|---|
| $d_{111}$=2.961Å | $d_{111}$=2.9516Å | $\varepsilon_{d_{111}} \sim -0.3175\%$ |
| $d_{\bar{1}11}$=2.961Å | $d_{\bar{1}11}$=2.953Å | $\varepsilon_{d_{\bar{1}11}} \sim -0.2702\%$ |
| $d_{0\bar{2}2}$=1.793Å | $d_{0\bar{2}2}$=1.808Å | $\varepsilon_{d_{0\bar{2}2}} \sim 0.8\%$ |
| $d_{\bar{2}11}$=2.121Å | $d_{\bar{2}11}$=2.0979 Å | $\varepsilon_{d_{\bar{2}11}} \sim -1\%$ |

**Table S1.** Lattice parameters of $ZrO_2$ thin films calculated using XRD and STEM data and the strains along different crystal orientations

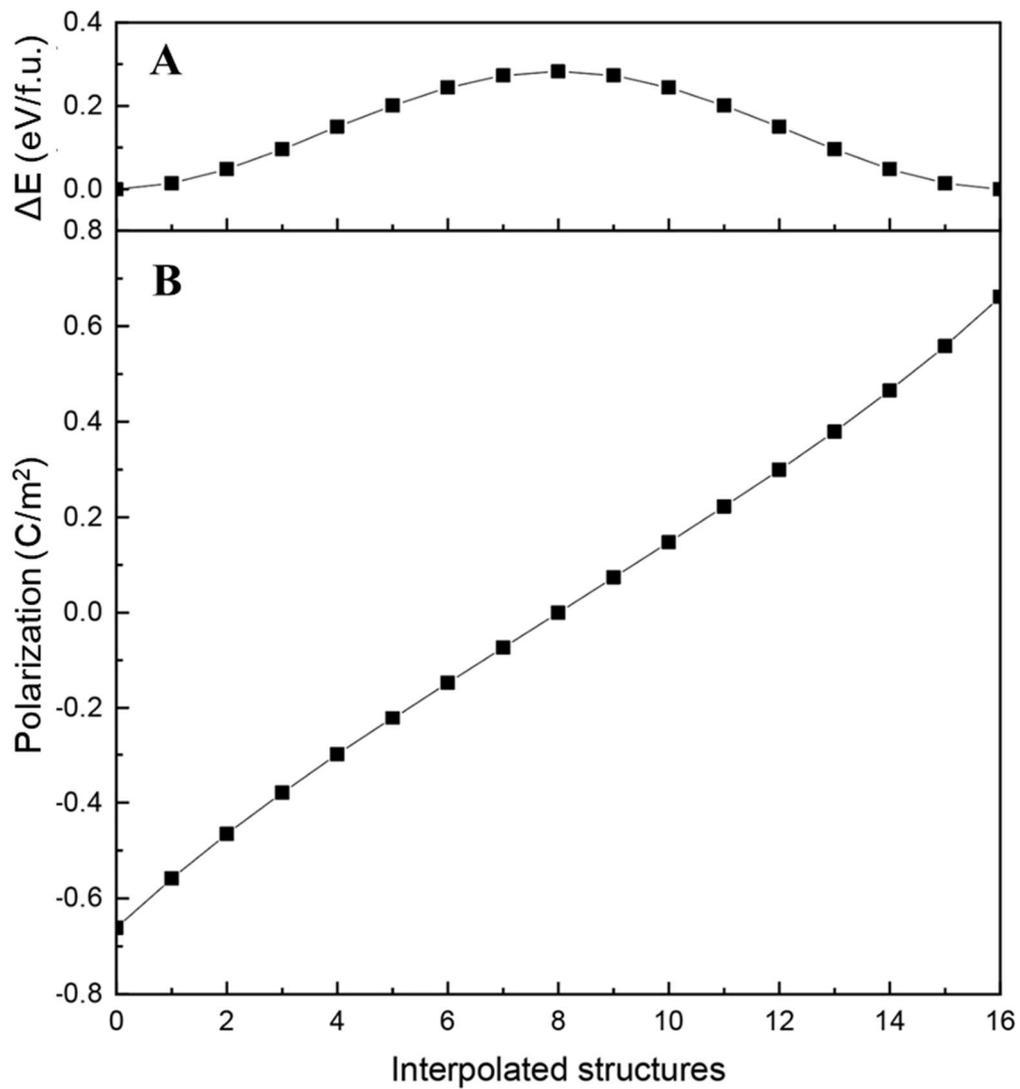

**Fig. S6. DFT computed evolution of ferroelectric polarization for the *o-FE* phase along the polarization flipping path and the corresponding energy profile.** (**A**) A set of structures was interpolated between the polarization up and down states in order to exclude the contribution from the artificial 'quanta' due to ambiguous definition of electric polarization in a periodic system. (**B**) The theoretical predicted FE polarization value is ~66 $\mu C/cm^2$.